\def\la{\leftarrow}

\def\ra{\rightarrow}

\def\bd{\noindent\bf}

\def\sbd{\vspace{8pt}\noindent\bf}

\documentstyle[fleqn,12pt]{article}
\begin{document}

\title{On Statman's Finite Completeness Theorem}
\author{Richard Statman \ \ Gilles Dowek}

\date{June 1, 1992}


\maketitle

\abstract{
We give a complete self-contained proof of {\it Statman's finite completeness 
theorem} and of a corollary of this theorem stating that the 
$\lambda$-definability conjecture implies the higher-order matching 
conjecture.}

\bigskip

The purpose of this note is to give a complete self-contained proof of
{\it Statman's finite completeness theorem} and of a corollary of this theorem
stating that the $\lambda$-definability conjecture implies the higher-order
matching conjecture. Both results are proved in \cite{Statman82} (theorem 2 
and 5). Although, since \cite{Statman82} assumes some familiarity 
with typed $\lambda$-calculus model theory and presents several results in 
short space, it may be not very accessible to readers not familiar with this
subject.

Section 1 gives the basic notations used in the paper. The reader not familiar
with simply typed $\lambda$-calculus should consult Hindley and Seldin
\cite{HinSel}. 
Section 2 presents standard models for simply typed $\lambda$-calculus, 
it is based on Henkin \cite{Henkin}. Section 3 presents the  
Completeness theorem, it is based on Friedman \cite{Friedman},
Plotkin \cite{Plotkin73} \cite{Plotkin80} and Statman \cite{Statman85}.
Section 4 presents the construction of a model for some equational theories.
Section 5 presents Statman's finite completeness theorem. Both section 4 and 5
are based on \cite{Statman82}. Section 6 presents the $\lambda$-definability
conjecture. The notion of $\lambda$-definability is taken from Plotkin
\cite{Plotkin73} \cite{Plotkin80}. This conjecture has been studied by Plotkin
and Statman.
At last section 7 presents the higher order matching conjecture and the 
proof that the  $\lambda$-definability conjecture implies the higher order 
matching conjecture. The decidability of higher order matching is conjectured
in Huet \cite{Huet76}, the equivalence of the higher order matching problem and
the higher order matching problem with closed terms if proved in
\cite{Statman81}  and the proof that the $\lambda$-definability conjecture 
implies the higher order matching conjecture is from \cite{Statman82}, this 
proof is also discussed in Wolfram \cite{Wolfram}.

\section{Typed $\lambda$-calculus}

\subsection{Types}

The set of types is defined by 
$$T = \iota~|~(T \ra T)$$
The notation $\alpha \ra \beta \ra \gamma$ is an abbreviation for
($\alpha \ra (\beta \ra \gamma))$.\\
Obviously a type can be written in a unique way 
$\alpha = \beta_{1} \ra ... \ra \beta_{n} \ra \iota$.\\
If $\alpha$ is a type, the {\it order} of $\alpha$ ($o(\alpha)$) is inductively
defined by 
$$o(\iota) = 1$$
and 
$$o(\alpha \ra \beta) = max\{1+o(\alpha), o(\beta)\}$$

\subsection{Terms}

A {\it context} is a set of pairs $\Gamma = \{<x_{i}, \alpha_{i}>\}$ where 
$x_{i}$ is a variable and $\alpha_{i}$ is a type, such that if
$<x,\alpha> \in \Gamma$ and $<x,\beta> \in \Gamma$ then $\alpha = \beta$.\\
The set of terms is defined by
$$t = x~|~(t~t)~|~\lambda x:T.t$$
The notation $(t~u~v)$ is an abbreviation for $((t~u)~v)$.\\
The judgement {\it the term $t$ has type $\alpha$ in the context $\Gamma$} 
($\Gamma \vdash t:\alpha$) is inductively defined by
\begin{itemize}
\item if $<x,\alpha> \in \Gamma$ then $\Gamma \vdash x:\alpha$,
\item if $\Gamma \vdash t:\alpha \ra \beta$ and $\Gamma \vdash u:\alpha$
then $\Gamma \vdash (t~u):\beta$,
\item if $\Gamma \cup \{x:\alpha\} \vdash t:\beta$ then 
$\Gamma \vdash \lambda x:\alpha.t:\alpha \ra \beta$.
\end{itemize}
Obviously if a term $t$ has type $\alpha$ and $\beta$ in a context $\Gamma$ 
then $\alpha = \beta$.\\
We write $\Lambda^{\Gamma}_{\alpha}$ for the set of terms $t$ such that
$\Gamma \vdash t:\alpha$.

\subsection{Normalization}

If $t$ and $u$ are terms, we write $t[x \la u]$ for the term
obtained by substituting the free occurrences of $x$ by $u$.
We write $t \rhd u$ when $t$, $\beta \eta$-reduces in some steps to $u$.
A term is said to be $\beta \eta$-normal if it does not contain any redexes.
We write $t =_{\beta \eta} u$ when $t$ and $u$ are $\beta \eta$-equivalent.
As proved in \cite{HinSel} the reduction relation is strongly normalizable and
confluent on well-typed terms, thus a well-typed term has a unique 
$\beta \eta$-normal form.\\
Obviously a normal term can be written in a unique way
$$t = \lambda x_{1}:\alpha_{1}. ... \lambda x_{n}:\alpha_{n}.
(x~c_{1}~...~c_{p})$$ where $x$ is a variable.\\
Let 
$$t = \lambda x_{1}:\alpha_{1}. ... \lambda x_{n}:\alpha_{n}.
(x~c_{1}~...~c_{p})$$
a normal term of type $\alpha_{1} \ra ... \ra \alpha_{m} \ra \iota$.
The normal $\eta$-long form of $t$ is inductively defined as
$$t' = \lambda x_{1}:\alpha_{1}. ... \lambda x_{n}:\alpha_{n}.
\lambda x_{n+1}:\alpha_{n+1}. ... \lambda x_{m}:\alpha_{m}.
(x~c'_{1}~...~c'_{p}~x'_{n+1}~...~x'_{m})$$
where $c'_{i}$ is the normal $\eta$-long form of $c_{i}$ and
$x'_{j}$ is the normal $\eta$-long form of $x_{j}$.

\section{Standard Models}

In typed $\lambda$-calculus model theory, we do not look at $\lambda$-terms
as at functions but rather as at {\it notations} for set-theoretical functions.

{\sbd Definition:} Standard Model

A {\it standard model} is a family of sets indexed by types 
$(M_{\alpha})_{\alpha}$ such that 
$M_{\alpha \ra \beta} = M_{\beta}^{M_{\alpha}}$.

{\sbd Definition:} Assignment 

Let $\Gamma$ be a context $(M_{\alpha})_{\alpha}$ be a standard model,
an {\it assignment from $\Gamma$ onto $(M_{\alpha})_{\alpha}$} is a function 
$\nu$ which maps every variable 
$x$ of type $\alpha$ of $\Gamma$ to an element of $M_{\alpha}$.

{\sbd Definition:} Interpretation

Let $\Gamma$ be a context, $(M_{\alpha})_{\alpha}$ be a standard model
and $\nu$ an assignment from this context onto this model, we define the 
function $\tilde{\nu}$ which maps every term of $\Lambda_{\alpha}^{\Gamma}$ to 
an element of $M_{\alpha}$ by
\begin{itemize}
\item $\tilde{\nu}(x) = \nu(x)$,
\item if $u$ is a term of type $\alpha \ra \beta$ and $v$ of type $\alpha$ then 
$\tilde{\nu}(u~v) = \tilde{\nu}(u)(\tilde{\nu}(v))$,
\item if $u$ is a term of type $\beta$ then for all $d \in M_{\alpha}$, 
$\tilde{\nu}(\lambda x:\alpha.u) (d) = \tilde{\nu^{+}}(u)$
where $\nu^{+}(x) = d$ and $\nu^{+}(y) = \nu(y)$ for $y \neq x$.

\end{itemize}

{\sbd Notation:} Let $(M_{\alpha})_{\alpha}$ be a standard model and 
$t$ and $u$ two terms with the same type $\alpha$, we write 
$(M_{\alpha})_{\alpha} \models t = u$ if for each assignment $\nu$,
we have $\tilde{\nu}(t) = \tilde{\nu}(u)$.

{\sbd Remark:} Given a set $M$, there exists only one standard model, 
$(M_{\alpha})_{\alpha}$ such that $M_{\iota} = M$.

Moreover if $M$ and $M'$ are two sets with the same cardinal, then 
we can construct an obvious isomorphism between the standard models 
$(M_{\alpha})_{\alpha}$ and $(M'_{\alpha})_{\alpha}$ based on
$M$ and $M'$. Indeed let us consider a bijection $\Phi$ between $M$ and $M'$ we
construct a family of bijections $\Phi_{\alpha}$ between $M_{\alpha}$ and
$M'_{\alpha}$ by $\Phi_{\iota} = \Phi$ and 
$\Phi_{\alpha \ra \beta} (f) = \Phi_{\beta} \circ f \circ \Phi_{\alpha}^{-1}$.

Obviously if $\nu$ is an assignment onto $(M_{\alpha})_{\alpha}$ then 
the function $\nu'$ which maps every variable $x$ of type $\alpha$ to 
the element $\Phi_{\alpha} (\nu(x))$ is an assignment onto 
$(M'_{\alpha})_{\alpha}$ and for each every $t$ of type $\alpha$
$\tilde{\nu'} (t) = \Phi_{\alpha}(\tilde{\nu}(t))$.
So if $t$ and $u$ are two terms of the same type then 
$(M_{\alpha})_{\alpha} \models t = u$ if and only if 
$(M'_{\alpha})_{\alpha} \models t = u$.

So given a cardinal $\xi$ there exists only one standard model (up to 
isomorphism) such that $M_{\iota}$ has cardinal $\xi$, we write it $M_{\xi}$.

\section{Completeness Theorem}

{\sbd Definition:} Friedman-Plotkin Model

Assume $\Gamma$ contains an infinite number of variables of each type.
Let $M_{\iota} = \Lambda_{\iota}^{\Gamma} /=_{\beta \eta}$ and 
$(M_{\alpha})_{\alpha}$ the standard model built on this set. 

{\sbd Proposition:}
There exists an assignment $\nu$ such that for every $t$ of type $\iota$
we have $\tilde{\nu}(t) = t /=_{\beta \eta}$. 

{\bd Proof:} 
We construct the assignment $\nu$ by induction over the order of the types of 
the variables of $\Gamma$. 
If $x$ has type $\iota$ then $\nu(x) = x /=_{\beta \eta}$. Then
assume the definition of $\nu(x)$ given for all the variables $x$ of order
strictly lower to $k$.
Let $\beta = \gamma_{1} \ra ... \ra \gamma_{n} \ra \iota$ be a type of order 
lower or equal to $k$, $t$ a term of type $\beta$ and $d \in M_{\beta}$, 
we write $t ~\tilde{\in}~ d$ if for all variables
$x_{1}:\gamma_{1}, ..., x_{n}:\gamma_{n}$ which do not occur free in $t$
we have $(t~x_{1}~...~x_{n}) \in d(\nu(x_{1})) ... (\nu(x_{n}))$.
Obviously if $t ~\tilde{\in}~ d$ and $u ~\tilde{\in}~ d$ then 
$t =_{\beta \eta} u$. 

Let $x$ be a variable of type
$\alpha = \beta_{1} \ra ... \ra \beta_{n} \ra \iota$ of order $k$, 
we define $\nu(x)$ by
$$\nu(x)(d_{1}) ... (d_{n}) = (x~t_{1}~...~t_{n}) /=_{\beta \eta}$$
if there exists $t_{1} ~\tilde{\in}~ d_{1}$, ...,  
$t_{n} ~\tilde{\in}~ d_{n}$ (obviously the element 
$(x~t_{1}~...~t_{n}) /=_{\beta \eta}$ does not depend of the choice
of $t_{1}, ...., t_{n}$) and $\nu(x)(d_{1}) ... (d_{n})$ be anything otherwise.

We prove by induction on the structure of the normal $\eta$-long form of
$t$ that $t ~\tilde{\in}~ \tilde{\nu}(t)$. 
Let $t$ be a term of type $\beta_{1} \ra ... \ra \beta_{n} \ra \iota$.
Since there is in $\Gamma$ an infinite number of variables of each type
there are in $\Gamma$ variables $x_{1}:\beta_{1}$, ..., $x_{n}:\beta_{n}$ 
which do not occur free in $t$. 
Modulo bound variable renaming the term $t$ can be written
$$t =
     \lambda x_{1}:\beta_{1}.... \lambda x_{n}:\beta_{n}. (x~u_{1}~...~u_{p})$$
By induction hypothesis, for every $i$, we have 
$$u_{i} ~\tilde{\in}~ \tilde{\nu}(u_{i})$$
so by definition of $\nu$ we have
$$\nu(x) (\tilde{\nu}(u_{1})) ... (\tilde{\nu}(u_{p})) = 
(x~u_{1}~...~u_{n}) /=_{\beta \eta}$$
i.e.
$$(x~u_{1}~...~u_{n}) \in 
\nu(x) (\tilde{\nu}(u_{1})) ... (\tilde{\nu}(u_{p}))$$
i.e.
$$(x~u_{1}~...~u_{n}) \in \tilde{\nu}(x~u_{1}~...~u_{p})$$
so\\
\hspace*{10pt}
$(\lambda x_{1}:\beta_{1}....\lambda x_{n}:\beta_{n}.(x~u_{1}~...~u_{p})
~x_{1}~...~x_{n})$

\hfill $\in \tilde{\nu}(\lambda x_{1}:\beta_{1}....\lambda x_{n}:\beta_{n}.
(x~u_{1}~...~u_{p})) (\nu(x_{1})) ... (\nu(x_{n}))$

\noindent i.e.
$$(t~x_{1}~...~x_{n}) \in \tilde{\nu}(t) (\nu(x_{1})) ... (\nu(x_{n}))$$
so
$$t ~\tilde{\in}~ \tilde{\nu}(t)$$
So if $t$ has type $\iota$ then
$t \in \tilde{\nu}(t)$, i.e. $\tilde{\nu}(t) = t /=_{\beta \eta}$. 

{\sbd Theorem:} (Friedman-Plotkin) Completeness Theorem

If $t$ and $u$ are terms of type $\alpha$ then
$(M_{\alpha})_{\alpha} \models t = u$ if and only if $t =_{\beta \eta} u$.

{\bd Proof:} 
Obviously, if $t =_{\beta \eta} u$ then 
$(M_{\alpha})_{\alpha} \models t = u$. 
Conversely, if we have $(M_{\alpha})_{\alpha} \models t = u$ then let us write
the type of $t$ and $u$ $\alpha = \beta_{1} \ra ... \ra \beta_{n} \ra \iota$.
Since there is in $\Gamma$ an infinite number of variables of each type, there
are in $\Gamma$ variables $x_{1}:\beta_{1}, ..., x_{n}:\beta_{n}$ which do not 
occur free in $t$ and $u$. We have 
$$(M_{\alpha})_{\alpha} \models (t~x_{1}~...~x_{n}) = (u~x_{1}~...~x_{n})$$
so using the previous proposition
$$(t~x_{1}~...~x_{n}) /=_{\beta \eta} = (u~x_{1}~...~x_{n}) /=_{\beta \eta}$$
i.e.
$$(t~x_{1}~...~x_{n}) =_{\beta \eta} (u~x_{1}~...~x_{n})$$
thus 
$$t =_{\beta \eta} u$$

{\sbd Corollary:}
Let $t$ and $u$ be two terms of type $\alpha$, $t =_{\beta \eta} u$ if and only
if for all standard models $M$, $M \models t = u$.

{\sbd Corollary:}
If $\xi$ is an infinite cardinal and $t$ and $u$ are terms of of type $\alpha$
then $M_{\xi} \models t = u$ if and only if $t =_{\beta \eta} u$.

{\bd Proof:} Let us consider a context $\Gamma$ that contains $\xi$ variables
in the type $\iota$ and a denumerable number of variables in the other types.
The model $M$ constructed above is such that $M \models t = u$ if and only if
$t =_{\beta \eta} u$. This model is isomorphic to the model $M_{\xi}$ so 
$M_{\xi} \models t = u$ if and only if $t =_{\beta \eta} u$.

{\sbd Remark:} If $n$ in a finite cardinal then the completeness theorem fails
for the model $M_{n}$ because if $M_{\iota}$ is finite then $M_{\alpha}$ is 
finite for every type $\alpha$ and if $(M_{\alpha})_{\alpha}$ verifies the 
completeness theorem then $M_{\iota \ra (\iota \ra \iota) \ra \iota}$ is 
infinite. Indeed call
$$\overline{p} = \lambda x:\iota. \lambda f:\iota \ra \iota.(f ... (f~x) ...)$$
if $p$ and $q$ are distinct integers then
$\overline{p} \neq_{\beta \eta} \overline{q}$ and so
$\tilde{\nu}(\overline{p}) \neq \tilde{\nu}(\overline{q})$ and 
$M_{\iota \ra (\iota \ra \iota) \ra \iota}$ is infinite.

So $M_{\xi}$ verifies the completeness theorem if and only if $\xi$ is 
infinite.

\section{Equational Theories}

Let $\Gamma$ be a context such that all the types of the variables in $\Gamma$
are of order at most two.
Let $E$ be a set of equations $\{a_{i} = b_{i}\}$ where $a_{i}$ and $b_{i}$ are
well-typed in $\Gamma$ and have type $\iota$ in $\Gamma$. We 
consider the smallest equivalence relation compatible with term structure that 
contains $=_{\beta \eta}$ and the equations of $E$. We write this relation 
$=_{\beta \eta E}$.

{\sbd Definition:} The relation $=_{\beta \eta E}$

The relation $=_{\beta \eta E}$ is inductivelly defined by:
\begin{itemize}
\item if $t =_{\beta \eta} u$ then $t =_{\beta \eta E} u$,
\item if $a = b \in E$ and $c$ is a term of type $\iota \ra \alpha$ 
then $(c~a) =_{\beta \eta E} (c~b)$,
\item if $t =_{\beta \eta E} u$ then $u =_{\beta \eta E} t$,
\item if $t =_{\beta \eta E} u$ and $ u =_{\beta \eta E} v$ then 
$t =_{\beta \eta E} v$,
\end{itemize}

{\sbd Definition:} Statman's Model 

Let $M_{\iota} = \Lambda^{\Gamma}_{\iota} /=_{\beta \eta E}$.
Let $(M_{\alpha})_{\alpha}$ be the standard model built on $M_{\iota}$.

{\sbd Proposition:}
If $(M_{\alpha})_{\alpha} \models t = u$ then $t =_{\beta \eta E} u$.

{\bd Proof:}
Let the assignment $\nu$ be defined defined as
\begin{itemize}
\item{if $x$ has type $\iota$, then $\nu(x) = x /=_{\beta \eta E}$,}
\item{if $x$ has type $\iota \ra ... \ra \iota \ra \iota$ then
$\nu(x) = f$ where 
$$f (d_{1}) ... (d_{n}) = (x~u_{1}~...~u_{n}) /=_{\beta \eta E}$$
with $u_{1} \in d_{1}$, ..., $u_{n} \in d_{n}$ (obviously the class of 
the term $(x~u_{1}~...~u_{n})$ does not depend on the choice of 
$u_{1}, ..., u_{n}$).}
\end{itemize}
By induction over the structure of the normal form of $t$, we have
for every term $t$ of type $\iota$, $\tilde{\nu}(t) = t /=_{\beta \eta E}$.
Indeed, because $t$ has type $\iota$ its normal form can be written
$t = (x~u_{1}~...~u_{n})$ since $t$ is well-typed in $\Gamma$ the variable
$x$ is of order at most two and the $u_{i}$ have type $\iota$. 
So $\tilde{\nu}(u_{1}) = u_{1} /=_{\beta \eta E}$, ..., 
$\tilde{\nu}(u_{n}) = u_{n} /=_{\beta \eta E}$.
And by definition of $\nu$, $\nu(x)$ maps $u_{1} /=_{\beta \eta E}$, ..., 
$u_{n} /=_{\beta \eta E}$ to $(x~u_{1}~...~u_{n}) /=_{\beta \eta E}$.
So $\tilde{\nu}(x~u_{1}~...~u_{n}) = (x~u_{1}~...~u_{n}) /=_{\beta \eta E}$.

If $(M_{\alpha})_{\alpha} \models t = u$ then 
$\tilde{\nu}(t) = \tilde{\nu}(u)$ so 
$t  /=_{\beta \eta E} = u  /=_{\beta \eta E}$ i.e. $t =_{\beta \eta E} u$.

{\sbd Remark:} The converse is obviously false. If $E$ contains an equation
$x = y$ where $x$ and $y$ are two variables of type $\iota$ and $M$ is a non 
trivial model then there exists an assignment $\nu$ such that 
$\nu(x) \neq \nu(y)$.

{\sbd Remark:} The proposition is obviously false if $t$ and $u$ do not have
type $\iota$. Indeed consider a variable $f$ of type 
$\iota \ra \iota$ and a set $E$ which contains the equations $(f~t) = t$ 
for all the terms $t$ of type $\iota$, we have 
$M \models f = \lambda x:\iota. x$ but 
$f \neq_{\beta \eta E} \lambda x:\iota. x$.

{\sbd Remark:} The proposition is obviously false if $\Gamma$ contains 
variables of order greater than two. Indeed consider a variable $f$ 
of type $\iota \ra \iota$ and $F$ of type $(\iota \ra \iota) \ra \iota$ and
a set $E$ which contains the equations $(f~t) = t$ for all the terms $t$ of 
type $\iota$, we have $M \models (F~f) = (F~\lambda x:\iota. x)$ 
but $(F~f) \neq_{\beta \eta E} (F~\lambda x:\iota. x)$.

\section{Finite Models}

{\sbd Definition:} A model is said to be {\it finite} if the set $M_{\iota}$ 
is finite.

\medskip

We want to sharpen the completeness theorem of section 3 and build a finite
model. As remarked in section 3, the completeness theorem fails for such a
model, so our completeness requirement will be weaker. For each closed 
term $t$, we are going to construct a finite model $M_{t}$ such that for
each closed term $u$, $M_{t} \models t = u$ if and only if 
$t =_{\beta \eta} u$. We do not require the model $M_{t}$ to be uniform 
over $t$. 

\subsection{A Remark on the Relation $=_{\beta \eta E}$}

{\sbd Proposition:} Let $E = \{a_{i} = b_{i}\}$ be a set of equations such that
for every $i$, $a_{i}$ and $b_{i}$ have type $\iota$.
Let $t$ and $u$ be two terms such that $t =_{\beta \eta E} u$. Then either 
the normal forms of $t$ and $u$ are identical or they both have a subterm in 
the set $T = \{a_{i}, b_{i}\}$.

{\bd Proof:} By induction on the structure of the proof of 
$t =_{\beta \eta E} u$.
\begin{itemize}
\item If $t =_{\beta \eta} u$ then the normal forms of $t$ and $u$ are 
identical.
\item If $t = (c~a)$ and $u = (c~b)$ then if $x$ has an occurrence in the 
normal form of $(c~x)$ then $a$ is a subterm of the normal form of $(c~a)$ and
$b$ is a subterm of the normal form of $(c~b)$. Otherwise the terms $(c~a)$
and $(c~b)$ have the same normal form.
\item if $u =_{\beta \eta E} t$ then by induction hypothesis, either the normal
forms of $u$ and $t$ are identical of they both have a subterm in $T$.
\item If $t =_{\beta \eta E} v$ and $v =_{\beta \eta E} u$ then call $t'$, 
$v'$, $u'$ the normal forms of $t$, $v$, $u$. By induction hypothesis, either
\begin{itemize}
\item $t' = v'$ and $v' = u'$, in this case $t' = u'$,
\item $t'$ and $v'$ have a subterm in $T$ and $v' = u'$, in this case $t'$ and 
$u'$ have a subterm in $T$,
\item $t' = v'$ and $v'$ and $u'$ have a subterm in $T$, in this case $t'$ and
$u'$ have a subterm in $T$,
\item $t'$ and $v'$ have a subterm in $T$ and $v'$ and $u'$ have a subterm in 
$T$, in this case $t'$ and $u'$ have a subterm in $T$.
\end{itemize}
\end{itemize}

{\sbd Corollary:} Let $E = \{a_{i} = b_{i}\}$ be a set of equations such that
for every $i$, $a_{i}$ and $b_{i}$ have type $\iota$.
Let $t$ and $u$ be two terms such that no subterm of the normal form of $t$
is an $a_{i}$ or a $b_{i}$, then $t =_{\beta \eta E} u$ if and only if
$t =_{\beta \eta} u$.

\subsection{Finite Models for Terms of Order Lower than Three}

Before giving the theorem in its full generality, we shall consider the 
simpler case in which the order of the type of $t$ is lower than three.

{\sbd Proposition:} Let $t$ be a closed term which type is of order at most 
three.
There exists a finite model $M_{t}$ such that $M_{t} \models t = u$ if and
only if $t =_{\beta \eta} u$, and the number of elements of $M_{\iota}$ is 
computable in function of $t$.

{\bd Proof:} Let $\alpha = \beta_{1} \ra ... \ra \beta_{n} \ra \iota$ be the type
of $t$. Let $\Gamma$ be the context 
$\Gamma = \{x_{1}:\beta_{1}, ..., x_{n}:\beta_{n}\}$. The types of
the variables of $\Gamma$ are of order at most two. Let $u$ be a closed term of
type $\alpha$, we have $t =_{\beta \eta} u$ if 
and only if $(t~x_{1}~...~x_{n}) =_{\beta \eta} (u~x_{1}~...~x_{n})$. Let $E$ 
be the set containing all the equations $a = b$ with $a$ and $b$ in 
$\Lambda_{\iota}^{\Gamma}$ and neither $a$ nor $b$ is a subterm of the normal
form of $(t~x_{1}~...~x_{n})$. Using the proposition above 
$(t~x_{1}~...~x_{n}) =_{\beta \eta} (u~x_{1}~...~x_{n})$ if and only if 
$(t~x_{1}~...~x_{n}) =_{\beta \eta E} (u~x_{1}~...~x_{n})$.

Let us consider the model $M_{t}$ constructed at the section 4. 
Obviously if $t =_{\beta \eta} u$ then $M_{t} \models t = u$.
Conversely if we have $M_{t} \models t = u$ then 
$M_{t} \models (t~x_{1}~...~x_{n}) = (u~x_{1}~...~x_{n})$.
So $(t~x_{1}~...~x_{n}) =_{\beta \eta E} (u~x_{1}~...~x_{n})$, therefore 
$(t~x_{1}~...~x_{n}) =_{\beta \eta} (u~x_{1}~...~x_{n})$ and 
$t =_{\beta \eta} u$.

The number of elements of $M_{\iota}$ is $1+k$ where $k$ is the number of
distinct subterms of the normal form of $(t~x_{1}~...~x_{n})$ of type $\iota$,
it is therefore a computable in function of $t$.

\subsection{General Case}

{\sbd Definition:} Length of a Term

Let 
$t = \lambda x_{1}:\alpha_{1}. ... \lambda x_{n}:\alpha_{n}. 
(x~d_{1}~...~d_{n})$ 
be a normal $\eta$-long term, we define the {\it length} of $t$ ($|t|$) 
by induction on the number of variables occurrences of $t$ by 
$$|t| = 
1 + max \{|\lambda x_{1}:\alpha_{1}. ... \lambda x_{n}:\alpha_{n}. d_{i}|\}$$

\medskip

Consider a closed term $t$ of type $\alpha$, we shall prove that we can find
terms $w_{1}, ..., w_{p}$ of type $\alpha \ra \iota$ which free variables are
of order at most two and such that for each closed term $u$ of type $\alpha$,
$t =_{\beta \eta} u$ if and only if for all $i$, 
$(w_{i}~t) =_{\beta \eta} (w_{i}~u)$. Then we will be able to conclude as 
above.

Without loss of generality, we can assume that $t$ is normal $\eta$-long. 
We shall first construct a term $w$ such that $(w~t) \neq_{\beta \eta} (w~u)$ 
for all the normal $\eta$-long closed terms $u$ that do not have the same 
length as $t$. Then for each normal $\eta$-long closed term $u$ such that 
$u \neq t$ and $u$ has the same length as $t$ we shall construct a term $w$ 
such that $(w~t) \neq_{\beta \eta} (w~u)$. Since the number of normal 
$\eta$-long closed terms which have a given length and a given type is finite,
this will give us a finite number of $w_{i}$.

{\sbd Proposition:}
Let $t$ be a normal $\eta$-long closed term of type $\alpha$, there exists a 
term $w$ of type $\alpha \ra \iota$ such that for every normal $\eta$-long 
closed term $u$ of type $\alpha$, if $|t| \neq |u|$ then 
$(w~t) \neq_{\beta \eta} (w~u)$ and the free variables of $w$ are of order at 
most two.

{\bd Proof:} For each integer $p$ consider a variable $z_{p}$ of type
$\iota \ra ... \ra \iota \ra \iota$. We define by induction over the structure 
of $\alpha$ a term $c_{\alpha}$.
Let us write 
$$\alpha = \beta_{1} \ra ... \ra \beta_{p} \ra \iota$$
and
$$\beta_{1} = \gamma_{1}^{1} \ra ... \ra \gamma_{q_{1}}^{1} \ra \iota$$
$$...$$
$$\beta_{p} = \gamma_{1}^{p} \ra ... \ra \gamma_{q_{p}}^{p} \ra \iota$$
Let us define
$$c_{\alpha} = \lambda x_{1}:\beta_{1}. ... \lambda x_{p}:\beta_{p}.
(z_{p}~(x_{1}~c_{\gamma_{1}^{1}} ... c_{\gamma_{q_{1}}^{1}})
~...~(x_{p}~c_{\gamma_{1}^{p}} ... c_{\gamma_{q_{p}}^{p}}))$$
Now consider a term
$$t = \lambda y_{1}:\alpha_{1}. ... \lambda y_{n}:\alpha_{n}.
(y_{i}~d_{1}~...~d_{p})$$
Let us write 
$$\alpha_{i} = \beta_{1} \ra ... \ra \beta_{p} \ra \iota$$
and
$$\beta_{1} = \gamma_{1}^{1} \ra ... \ra \gamma_{q_{1}}^{1} \ra \iota$$
$$...$$
$$\beta_{p} = \gamma_{1}^{p} \ra ... \ra \gamma_{q_{p}}^{p} \ra \iota$$
Let us define
$$e_{1} = \lambda y_{1}:\alpha_{1}. ... \lambda y_{n}:\alpha_{n}.d_{1}$$
$$...$$
$$e_{p} = \lambda y_{1}:\alpha_{1}. ... \lambda y_{n}:\alpha_{n}.d_{p}$$
We have 
$$(t~c_{\alpha_{1}}~...~c_{\alpha_{n}}) =_{\beta \eta}
(z_{p}~
(e_{1}~c_{\alpha_{1}}~...~c_{\alpha_{n}}
~c_{\gamma_{1}^{1}}~...~c_{\gamma_{m_{1}}^{1}})
~...~
(e_{p}~c_{\alpha_{1}}~...~c_{\alpha_{n}}
~c_{\gamma_{1}^{p}}~...~c_{\gamma_{m_{p}}^{p}}))$$
By induction on the length of $t$, the length of the normal 
form of the term $(t~c_{\alpha_{1}}~...~c_{\alpha_{n}})$ is the length of $t$.
So if $|t| \neq |u|$ then the length of the normal forms of
the terms $(t~c_{\alpha_{1}}~...~c_{\alpha_{n}})$ and 
$(u~c_{\alpha_{1}}~...~c_{\alpha_{n}})$ are different and
$(t~c_{\alpha_{1}}~...~c_{\alpha_{n}}) 
\neq_{\beta \eta} (u~c_{\alpha_{1}}~...~c_{\alpha_{n}})$.

We take $w = \lambda x:\alpha.(x~c_{\alpha_{1}}~...~c_{\alpha_{n}})$.
If $|t| \neq |u|$ then $(w~t) \neq_{\beta \eta} (w~u)$.

{\sbd Proposition:} 
Let $t$ and $u$ be two distinct normal $\eta$-long closed terms of type 
$\alpha = \alpha_{1} \ra ... \ra \alpha_{n} \ra \iota$. There are terms 
$c_{1}, ...,  c_{n}$ of type
$\alpha_{1}, ..., \alpha_{n}$ such that 
$(t~c_{1}~...~c_{n}) \neq_{\beta \eta} (u~c_{1}~...~c_{n})$ and the free 
variables of $c_{1}, ...,  c_{n}$ are of order at most two.

{\bd Proof:} By induction on the number of variables occurrences of $t$.
Let us write 
$$t = \lambda x_{1}:\alpha_{1}. ... \lambda x_{n}:\alpha_{n}. 
(x_{i}~d_{1}~...~d_{p})$$
and 
$$u = \lambda x_{1}:\alpha_{1}. ... \lambda x_{n}:\alpha_{n}.
(x_{i'}~d'_{1}~...~d'_{p'})$$

{\it First Case:} $i \neq i'$ 

Let $\alpha_{i} = \beta_{1} \ra ... \ra \beta_{p} \ra \iota$ and 
$\alpha_{i'} = \beta'_{1} \ra ... \ra \beta'_{p'} \ra \iota$. We let $z$ and
$z'$ be two new variables of type $\iota$ and we take 
$c_{i} = \lambda y_{1}:\beta_{1}. ... \lambda y_{p}:\beta_{p}. z$ and
$c_{i'} = \lambda y_{1}:\beta'_{1}. ... \lambda y_{p'}:\beta'_{p'}. z'$ and
$c_{j}$ be any term when $j \neq i$ and $j \neq i'$.
We have $(t~c_{1}~...~c_{n}) =_{\beta \eta} z$
and $(u~c_{1}~...~c_{n}) =_{\beta \eta} z'$
so $(t~c_{1}~...~c_{n}) \neq_{\beta \eta} (u~c_{1}~...~c_{n})$.

{\it Second Case:} $i = i'$

So we have $p = p'$.
For all $j$ we let
$e_{j} =  \lambda x_{1}:\alpha_{1}. ... \lambda x_{n}:\alpha_{n}. d_{j}$
and
$e'_{j} =  \lambda x_{1}:\alpha_{1}. ... \lambda x_{n}:\alpha_{n}. d'_{j}$.
We have
$$t =_{\beta \eta} \lambda x_{1}:\alpha_{1}. ... \lambda x_{n}:\alpha_{n}. 
(x_{i}~(e_{1}~x_{1}~...~x_{n})~...~(e_{p}~x_{1}~...~x_{n}))$$
and 
$$u =_{\beta \eta} \lambda x_{1}:\alpha_{1}. ... \lambda x_{n}:\alpha_{n}. 
(x_{i}~(e'_{1}~x_{1}~...~x_{n})~...~(e'_{p}~x_{1}~...~x_{n}))$$
Since $t \neq_{\beta \eta} u$ there exists an integer $k$ such that 
$e_{k} \neq_{\beta \eta} e'_{k}$.
Let us write $\alpha_{i} = \beta_{1} \ra ... \ra \beta_{p} \ra \iota$
and $\beta_{k} = \gamma_{1} \ra ... \ra \gamma_{m} \ra \iota$.
The type of the terms $e_{k}$ and $e'_{k}$ is 
$\alpha_{1} \ra ... \ra \alpha_{n} \ra \gamma_{1} \ra ... \ra \gamma_{m}
\ra \iota$.
By induction hypothesis there exists terms 
$a_{1}, ..., a_{n}, b_{1}, ..., b_{m}$ such that 
$$(e_{k}~a_{1}~...~a_{n}~b_{1}~...~b_{m}) \neq_{\beta \eta} 
(e'_{k}~a_{1}~...~a_{n}~b_{1}~...~b_{m})$$
We let $z$ be a new variable of type $\iota \ra \iota \ra \iota$. We let
$$c_{i} = \lambda y_{1}:\beta_{1}. ...  \lambda y_{p}:\beta_{p}.
(z~(y_{k}~b_{1}~...~b_{m})~(a_{i}~y_{1}~...~y_{p}))$$
and $c_{j} = a_{j}$ for every $j \neq i$.

Remark that for each $j$,
$c_{j} [z \la \lambda x:\iota. \lambda y:\iota.y] =_{\beta \eta} a_{j}$, so
$$(e_{k}~c_{1}~...~c_{n}~b_{1}~...~b_{m}) \neq_{\beta \eta}
(e'_{k}~c_{1}~...~c_{n}~b_{1}~...~b_{m})$$
otherwise we would have 
$(e_{k}~a_{1}~...~a_{n}~b_{1}~...~b_{m}) =_{\beta \eta}
(e'_{k}~a_{1}~...~a_{n}~b_{1}~...~b_{m})$
by substituting the variable $z$ by the term 
$\lambda x:\iota. \lambda y:\iota.y$.
Now 
$$(t~c_{1}~...~c_{n}) =_{\beta \eta} (z~(e_{k}~c_{1}~...~c_{n}~b_{1}~...~b_{m})
~(a_{i}~(e_{1}~c_{1}~...~c_{n})~...~(e_{p}~c_{1}~...~c_{n})))$$
and
$$(u~c_{1}~...~c_{n}) =_{\beta \eta} 
(z~(e'_{k}~c_{1}~...~c_{n}~b_{1}~...~b_{m})~
(a_{i}~(e'_{1}~c_{1}~...~c_{n})~...~(e'_{p}~c_{1}~...~c_{n})))$$
So  
$$(t~c_{1}~...~c_{n}) \neq_{\beta \eta} (u~c_{1}~...~c_{n})$$

{\sbd Proposition:}
Let $t$ and $u$ be two distinct normal $\eta$-long closed terms of type 
$\alpha$, there exists a term $w$ of type $\alpha \ra \iota$ such that 
$(w~t) \neq_{\beta \eta} (w~u)$ and the free variables of $w$ are of order at 
most two.

{\bd Proof:}
Let $\alpha = \alpha_{1} \ra ... \ra \alpha_{n} \ra \iota$, by the previous
proposition there are terms $c_{1}, ..., c_{n}$ such that
$(t~c_{1}~...~c_{n}) \neq_{\beta \eta} (u~c_{1}~...~c_{n})$, and the free 
variables of $c_{1}, ..., c_{n}$ are of order at most two. We take
$w = \lambda x:\alpha .(x~c_{1}~...~c_{n})$.

{\sbd Proposition:}
Let $t$ be a closed term $t$ of type $\alpha$, there exist terms 
$w_{1}, ..., w_{p}$ of type $\alpha \ra \iota$ which free variables are
of order at most two and such that for each term $u$, $t =_{\beta \eta} u$ if
and only if for all $i$, $(w_{i}~t) =_{\beta \eta} (w_{i}~u)$.

{\bd Proof:} 
We construct, using a previous proposition, a term $w$ such that for each
normal $\eta$-long closed term $u$ of type $\alpha$ which length is different 
from the length of the normal $\eta$-long form of $t$, 
$(w~t) \neq_{\beta \eta} (w~u)$.
Then for each normal $\eta$-long closed term $u$ of type $\alpha$ which has 
the same length as the normal $\eta$-long form of $t$ and which is different 
from the normal form of $t$, we construct, using a previous proposition, a 
term $w$ such that $(w~t) \neq_{\beta \eta} (w~u)$.
Since the number of normal $\eta$-long closed terms which have the same length
and the same type as the normal $\eta$-long form of $t$ is finite, this gives 
us a finite number of $w_{i}$.

Obviously, for each closed term $u$ of type $\alpha$, $t =_{\beta \eta} u$ if 
and only if for each integer $i$, $(w_{i}~t) =_{\beta \eta} (w_{i}~u)$.

{\sbd Theorem:} (Statman) Finite Completeness Theorem

Let $t$ be a closed term of type $\alpha$.
There exists a finite model $M_{t}$ such that $M_{t} \models t = u$ if 
and only if $t =_{\beta \eta} u$, and the number of elements of $M_{\iota}$ is 
computable in function of $t$.

{\bd Proof:} Let $w_{1}, ..., w_{n}$ the terms given by the propsition above.
Let $E$ be the set containing all the equations $a = b$ with $a$ and $b$ in 
$\Lambda_{\iota}^{\Gamma}$ and neither $a$ nor $b$ is a subterm of 
the normal form of an $(w_{i}~t)$ for some $i$.
Using a proposition above $(w_{i}~t) =_{\beta \eta} (w_{i}~u)$ if and only if 
$(w_{i}~t) =_{\beta \eta E} (w_{i}~u)$.

Let us consider the model $M_{t}$ constructed at the section 4. 
Obviously if $t =_{\beta \eta} u$ then we have $M_{t} \models t = u$.
Conversely if $M_{t} \models t = u$ then $M_{t} \models (w_{i}~t) = (w_{i}~u)$.
So $(w_{i}~t) =_{\beta \eta E} (w_{i}~u)$, therefore 
$(w_{i}~t) =_{\beta \eta} (w_{i}~u)$ and $t =_{\beta \eta} u$.

The number of elements of $M_{\iota}$ is $1+k$ where $k$ is the number of
distinct subterms of the normal form of $(w_{i}~t)$ of type $\iota$, since the 
terms $w_{i}$ are computable in function of $t$, the number of elements of 
$M_{\iota}$ is computable in function of $t$.

{\sbd Corollary:}
Let $t$ and $u$ two terms of type $\alpha$, $t =_{\beta \eta} u$ if and only
if for all the finite standard models $M$, $M \models t = u$.

\section{The $\lambda$-definability Conjecture}

Because a simple function as the identity over integers is an infinite 
set (although it has a finite description), set-theoretical functions are not 
usually computational objects. In the same way, because completeness theorems
concern usually infinite sets, model checking is not usually an effective 
decision procedure. Both argument fail when the sets involved are finite. 
Indeed, finite set-theoretical functions are computational objects 
(e.g. association lists) and finite model checking is an effective decision 
procedure (e.g. propositional calculus).

{\sbd Definition:}
Let $(M_{\alpha})_{\alpha}$ be the standard model with the base set 
$M = M_{\iota}$.
A function $f$ of $M_{\alpha}$ is said to be $\lambda$-definable if there 
exists a closed $\lambda$-term $t$ of type $\alpha$ such that 
$\tilde{\nu}(t) = f$ (where $\nu$ is the only assignment over the empty set).

{\sbd Conjecture:} $\lambda$-definability Conjecture

If $M$ is finite then it is decidable whether of not a function $f$ of 
$M_{\alpha}$ is $\lambda$-definable.

{\sbd Remark:} The problem makes sense because $M$ is finite, otherwise
the functions of $M_{\alpha}$ would not be computational objects.

\section{The $\lambda$-definability Conjecture Implies the
Higher Order Matching Conjecture}

{\sbd Definition:} Higher Order Matching Problem

A {\it higher order matching problem} is a pair of terms $<a,b>$ of
types $\alpha_{1} \ra ... \ra \alpha_{n} \ra \beta$ and $\beta$.
A solution to this problem is a n-uple of terms 
$<t_{1}, ..., t_{n}>$ of type 
$\alpha_{1}, ... , \alpha_{n}$ such that 
$(a~t_{1}~...~t_{n}) =_{\beta \eta} b$.

{\sbd Conjecture :} Higher Order Matching Conjecture

It is decidable whether of not a higher order matching problem has a solution.

{\sbd Proposition:} The higher order matching problem is decidable if and only
if the higher order matching problem with closed $a$ and $b$ is decidable.

{\bd Proof:} 
Let $<a,b>$ be a problem. Let $x_{1}:\beta_{1}, ..., x_{p}:\beta_{p}$
be the free variables occurring in $a$ and $b$. Let
$$\gamma_{1} = \beta_{1} \ra ... \ra \beta_{p} \ra \alpha_{1}$$
$$...$$
$$\gamma_{n} = \beta_{1} \ra ... \ra \beta_{p} \ra \alpha_{n}$$
$$a' = \lambda y_{1}:\gamma_{1}. ... \lambda y_{n}:\gamma_{n}.
       \lambda x_{1}:\beta_{1}. ... \lambda x_{p}:\beta_{p}.
       (a~(y_{1}~x_{1}~...~x_{p})~...~(y_{n}~x_{1}~...~x_{p}))$$
$$b' = \lambda x_{1}:\beta_{1}. ... \lambda x_{p}:\beta_{p}.b$$
If we have 
$$(a~t_{1}~...~t_{n}) = b$$
then 
$$(a'~
\lambda x_{1}:\beta_{1}. ... \lambda x_{p}:\beta_{p}.t_{1}~...~
\lambda x_{1}:\beta_{1}. ... \lambda x_{p}:\beta_{p}.t_{n}) = b'$$
Conversely if we have
$$(a'~t'_{1}~...~t'_{n}) = b'$$
then
$$(a~(t'_{1}~x_{1}~...~x_{p})~...~(t'_{n}~x_{1}~...~x_{p})) = b$$
So the problem $<a,b>$ has a solution if and only if the problem $<a',b'>$ 
has one.

{\sbd Proposition:} The higher order matching problem is decidable if and only
if the higher order matching problem with closed $t_{1}, ..., t_{n}$ is 
decidable.

{\bd Proof:}
Let $<a,b>$ be a problem. Let
$$a' = 
\lambda y_{1}:\iota \ra \alpha_{1}. ... \lambda y_{n}:\iota \ra \alpha_{n}.
\lambda x:\iota. (a~(y_{1}~x)~...~(y_{n}~x))$$
and 
$$b' = \lambda x:\iota.b$$
(remark that if $a$ and $b$ are closed then $a'$ and $b'$ also are).

Assume the problem $<a,b>$ has a solution $t_{1}, ..., t_{n}$.
Let $z$ be a variable of type $\iota$.
For each type $\beta = \gamma_{1} \ra ... \ra \gamma_{k} \ra \iota$
we consider the term 
$w_{\beta} = \lambda z_{1}:\gamma_{1}. ... \lambda z_{k}:\gamma_{k}.z$.

Let $x_{1}:\beta_{1}, ..., x_{p}:\beta_{p}$ be the variables occurring free
in the terms $t_{1}, ..., t_{n}$,
$t'_{i} = t_{i}[x_{1} \la w_{\beta_{1}}, ..., x_{k} \la w_{\beta_{k}}]$
and $u_{i} = \lambda z:\iota. t'_{i}$.
We have 
$$(a'~u_{1}~...~u_{n}) = b'$$
So the problem $<a',b'>$ has closed solution.

Now if the problem $<a',b'>$ has a closed solution $u_{1}, ..., u_{n}$ then 
let $z$ be a variable of type $\iota$ and $t_{i} = (u_{i}~z)$. We have 
$$(a'~u_{1}~...~u_{n}) = b'$$
so
$$\lambda x:\iota.(a~(u_{1}~x)~...~(u_{n}~x)) = \lambda x:\iota.b$$
so
$$(a~(u_{1}~z)~...~(u_{n}~z)) = b$$
i.e.
$$(a~t_{1}~...~t_{n}) = b$$

So the problem $<a,b>$ has a solution if and only if the problem $<a',b'>$ has 
a closed solution.

{\sbd Theorem:} (Statman) If $\lambda$-definability is decidable then higher 
order matching is decidable.

{\bd Proof:} Let us assume the $\lambda$-definability conjecture. We take 
two closed terms $a:\alpha_{1} \ra ... \alpha_{n} \ra \beta$ and $b:\beta$ and
consider the model $M_{b}$ constructed in section 5. Let $A = \tilde{\nu}(a)$ 
and $B = \tilde{\nu}(b)$ (where $\nu$ is the only assignment over the empty 
set).
By an enumeration procedure we select all the n-uples $<T_{1}, ..., T_{n}>$ 
such that $A(T_{1}, ..., T_{n}) = B$. The problem $<a,b>$ has a closed solution
if and only if there is such a n-uple such that all the $T_{i}$ are 
$\lambda$-definable.

Indeed if the problem $<a,b>$ has a closed solution $<t_{1}, ..., t_{n}>$, then
let $T_{i} = \tilde{\nu}(t_{i})$, the $T_{i}$ are $\lambda$-definable and
$A(T_{1}, ..., T_{n}) = B$.
Conversely if there are $\lambda$-definable $T_{1}, ...T_{n}$ such that
$A(T_{1}, ..., T_{n}) = B$. Let $t_{1}, ..., t_{n}$ be the terms such that
$T_{i} = \tilde{\nu}(t_{i})$. We have 
$M_{b} \models (a~t_{1}~...~t_{n}) = b$. 
So $(a~t_{1}~...~t_{n}) =_{\beta \eta} b$.

\end{document}